\documentclass[showpacs,preprint,prl]{revtex4}
\usepackage{graphicx}
\usepackage{color,ulem}

\definecolor{darkred}{rgb}{0.90,0,0}
\definecolor{darkgreen}{rgb}{0,0.60,.2}
\definecolor{darkblue}{rgb}{0,0,1}
\definecolor{grey}{cmyk}{0,0,0,0.25}
\definecolor{orange}{cmyk}{0,0.6,0.8,0}

\begin{document}

\title{Gate-tunable split Kondo Effect in a Carbon Nanotube Quantum Dot}
\author{A. Eichler
\footnote{present address: CIN2(ICN-CSIC), Catalan Institute of
Nanotechnology, Campus de la UAB 08193 Bellaterra (Barcelona)
Spain}}
\author{M. Weiss}\email{Markus.Weiss@unibas.ch}
\author{C. Sch\"onenberger}
\affiliation{Department of Physics, University of Basel,
Klingelbergstrasse 82, CH-4056 Basel, Switzerland}

\begin{abstract}
We show a detailed investigation of the split Kondo effect in a
Carbon nanotube quantum dot with multiple gate electrodes. Two
conductance peaks, observed at finite bias in nonlinear transport
measurements are found to approach each other for increasing
magnetic field, to result in a recovered zero-bias Kondo resonance
at finite magnetic field. Surprisingly, in the same charge state,
but under different gate-configurations, the splitting does not
disappear for any value of the magnetic field, but we observe an
avoided crossing. We think that our observations can be understood
in terms of a two-impurity Kondo effect with two spins coupled
antiferromagnetically. The exchange coupling between the two spins
can be influenced by a local gate, and the non-recovery of the Kondo
resonance for certain gate configurations is explained by the
existence of a small antisymmetric contribution to the exchange
interaction between the two spins.
\end{abstract}
\pacs{72.15.Qm,73.63.Kv,73.21.La,73.23.Hk,73.63.Fg}

\maketitle

The Kondo effect, the screening of a localized magnetic impurity by
a surrounding Fermi gas, is one of the most prominent examples of
correlated electron systems \cite{Hewson93}. The observation of the
Kondo effect in quantum dots has sparked renewed interest in the
exploration of this correlation effect, as quantum dots are highly
tunable, and a lot of different parameter regimes of a Kondo state
can be realized \cite{Goldhaber98, Cronenwett98}. Single-wall Carbon
nanotubes can be used for the production of quantum dot devices, and
allow for the study of the Kondo effect, as they usually show a
strong confinement potential, a fact that leads to a large Kondo
energy scale $k_B T_K$ \cite{Nygard00}. Carbon nanotubes can
moreover be contacted by superconducting \cite{Morpurgo99} and
ferromagnetic \cite{Tsukagoshi99, Sahoo05} electrodes, and therefore
allow the production of hybrid devices incorporating non-normal
metal contacts. One aspect in which Carbon nanotube quantum dots are
less versatile than e. g. lateral quantum dots, that are also widely
used to study the Kondo effect, is the control over the contact
transparencies and their asymmetry, two key parameters for the Kondo
effect. The total transparency is usually given after production of
a CNT quantum dot, and can only be changed in a small range by
driving the backgate over a large voltage. The contact asymmetry can
vary slightly between different charge states \cite{Eichler09}, but
up to now cannot be controlled systematically.

Here, we report on an experiment performed on a Carbon nanotube
quantum dot with two additional, local gate electrodes. The local
gates, fabricated as topgates on a thin, high-$\kappa$ oxide layer
were placed above the interface between the metal contact and the
Carbon nanotube, in an attempt to localize their gating effect to
the area close to the metal-nanotube interface. The coupling of the
quantum dot to the leads is in the intermediate regime, where
Coulomb interactions and Kondo correlations dominate the electronic
transport. We observe a split Kondo resonance, with a zero field
splitting of $\Delta E \approx$ 120~$\mu$eV. As we will point out
below, this observation is consistent with a two-impurity Kondo
effect, with two electrons coupled to each other
antiferromagnetically. In this sense, our experimental situation is
similar to previous experiments on lateral quantum dots
\cite{Craig04}, although in our case, the two-spin configuration is
more accidental.

\begin{figure}
\begin{center}
\includegraphics[width=10cm]{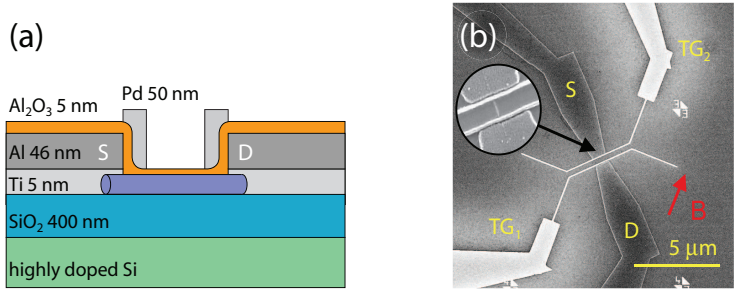}
\caption{ Cross section (a) and scanning electron micrograph (b) of
the nanotube with superconducting Ti/Al source (S) and drain (D)
electrodes, palladium topgates (TG$_1$, TG$_2$) with 5 nm of
Al$_2$O$_3$ topgate dielectric, and the highly doped silicon
substrate with 400~nm of thermal oxide, acting as backgate.
The red arrow in b) indicates the direction of the magnetic
field.\label{fig1}}
\end{center}
\end{figure}

A device schematic together with a scanning electron micrograph of a
device equivalent to the one that was measured is presented in
figure \ref{fig1}. Single-wall Carbon nanotubes are grown on
thermally oxidized silicon wafers by the chemical vapor deposition
method. We select individual, straight nanotubes with a
scanning electron microscope and define source (S) and drain (D)
electrodes by e-beam lithography. A thin (5 nm) layer of Al$_2$O$_3$
is then grown over the whole device by atomic layer deposition (ALD)
in a Cambridge Nanotech Savannah ALD system. In a last lithography
step palladium topgates are defined. A backgate voltage $V_{BG}$ is
applied to the highly doped silicon substrate. The sample is glued
into a nonmagnetic chip carrier and wire-bonded. We cool the sample
in a dilution refrigerator to a temperature of $20$~mK and measure
the differential conductance $G = dI/dV$ through the device with a
lock-in technique under a constant AC voltage bias of $5~\mu$V. The
aluminum contacts are superconducting below $\sim 1$~K, but we will
in the following concentrate on electrical transport under a
magnetic field $B$, that is applied in the plane of the wafer,
under an angle of about $45^{o}$ to the nanotube axis as indicated
in figure \ref{fig1}b. The effect we discuss emerges for $|B|
> 0.2$~T, when superconductivity is suppressed and the leads behave
as normal metals. $G$ is measured as a function of magnetic field
$B$, voltage $V$, backgate $V_{BG}$ and the voltage on one topgate
$V_{TG}$, with the second topgate floating. The backgate is expected
to affect the whole device, whereas the topgate should gate more
locally, in the vicinity of the CNT-metal contact interface.
Large range gate voltage sweeps revealed, that the nanotube in the
device presented here was metallic.

\begin{figure}
\begin{center}
\includegraphics[width=10cm]{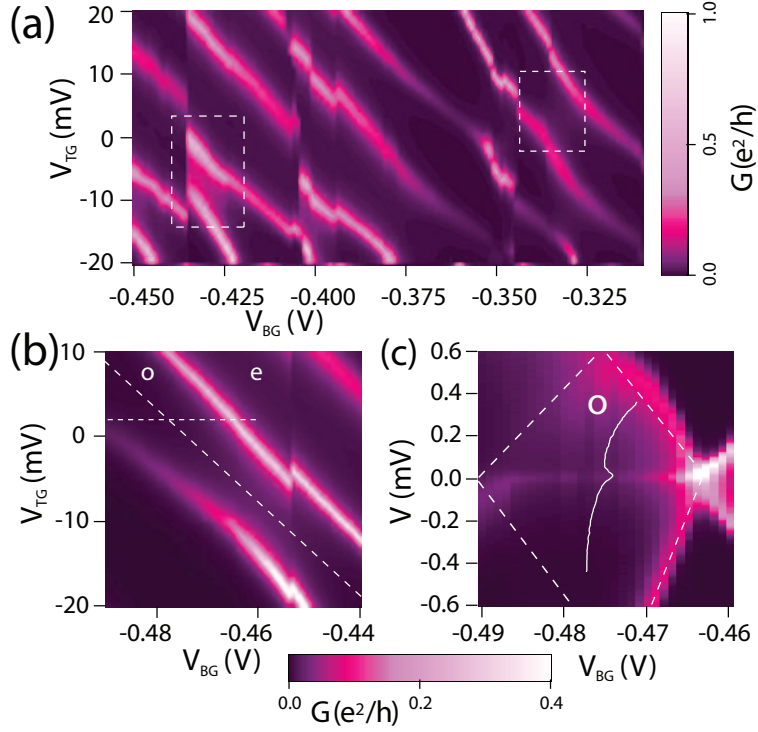}
\caption{a)~Plot of $G$ versus $V_{BG}$ and $V_{TG}$ for $V_{SD} =
0$~V and $B = 0.4$~T. Dashed frames indicate regions where two
resonances approach and the differential conductance between them
increases. b)~Plot of the region that is analyzed in detail ($B =
1.6$~T). The horizontal line marks the position of the measurement
in figure~\ref{fig2}c, and the slanted line of the one in
figure~\ref{fig3}. c)~$G$ versus $V_{BG}$ and $V$ for $V_{TG} =
-2$~mV and $B = 0.4$~T. A Kondo ridge is visible around $V = 0$
throughout the Coulomb diamond labeled \textbf{o}. \label{fig2}}
\end{center}
\end{figure}

Figure~\ref{fig2}a shows a measurement of $G$ as a function of
$V_{BG}$ and $V_{TG}$ at zero bias voltage ($V = 0$) and for $B =
0.4$~T. We can clearly distinguish resonances in $G(V_{BG},
V_{TG})$, corresponding to positions where an electronic level on
the quantum dot is energetically aligned with the chemical potential
of the contacts. The values of $V_{BG}$ for which this occurs shift
with $V_{TG}$. The slope of the resulting lines in the colorscale
plot of $G(V_{BG}, V_{TG})$ in figure~\ref{fig2}a is not constant,
however, but we observe regions where two resonances approach and
the differential conductance between them increases (dashed frames).
We shall see that this can be attributed to a Kondo resonance whose
energy and characteristics can be tuned by $V_{TG}$. Charge
rearrangements leading to jumps in the electrostatic potential of
the quantum dot can be seen as discontinuities in the colorscale
plot of $G(V_{BG}, V_{TG})$ in figure~\ref{fig2}a. As these jumps
occur more frequently in the present device compared to devices
without topgates, we assume that they are due to trapped charge in
the topgate oxide. The jumps do not occur frequent enough however,
to prohibit a careful study of the device.

For a detailed analysis, we concentrate on the ranges of $V_{BG}$
and $V_{TG}$ plotted in figure~\ref{fig2}b. We denote two charge
states by \textbf{o} (odd) and \textbf{e} (even) in anticipation of
a spin-$1/2$ Kondo effect in \textbf{o}. The horizontal dashed line
marks the position of the measurement of $G$ versus $V_{BG}$ and $V$
displayed in figure~\ref{fig2}c. From the size of the Coulomb
diamond in this plot, we extract a charging energy $U_C$ of about
$0.6$~meV (white dashed lines). In addition, a ridge of high
differential conductance shows up around zero bias, which we
identify as signature of the spin-$1/2$ Kondo effect.

\begin{figure}
\begin{center}
\includegraphics[width=10cm]{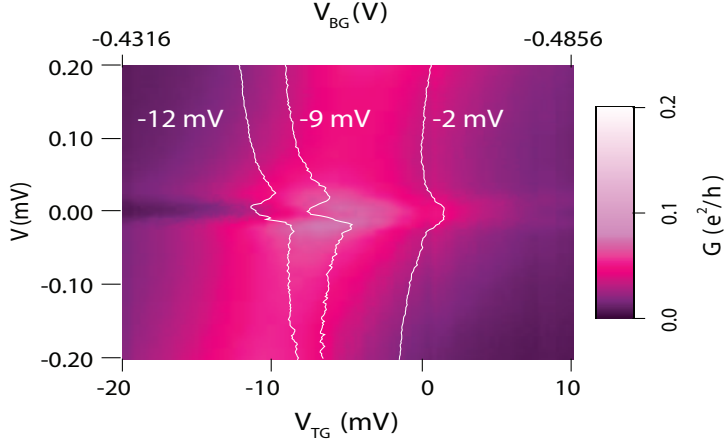}
\caption{$G(V)$ along the slanted dashed line in figure~\ref{fig2}b
at $B = 0.4$~T. $V_{BG}$ and $V_{TG}$ are driven simultaneously so
that the dot chemical potential stays constant in the middle of the
charge state. The Kondo peak at $V_{TG} = -2$~mV evolves into a dip
at $V_{TG} \leq -9$~mV. White lines correspond to cuts at the
voltage values used for the measurements in figure~\ref{fig4}.
\label{fig3}}
\end{center}
\end{figure}

Changing both $V_{BG}$ and $V_{TG}$ in the right way, we can modify
the electrostatic confinement potential of the dot while keeping the
chemical potential and the occupation constant. In figure~\ref{fig3}
we show a measurement of the nonlinear conductance $G(V)$ for a
range of gate voltages that are indicated by the slanted dashed line
in figure~\ref{fig2}b and that correspond to the middle of the
charge state denoted as {\bf o}. We can follow the development of
the Kondo feature from a zero bias peak ($V_{TG} = -2$~mV) into two
peaks at finite voltage ($V_{TG} \leq -9$~mV). The width of the peak
at $V_{TG} = -2$~mV is of the order of $50~\mu$V, corresponding to a
Kondo temperature of about $0.5$~K.

\begin{figure}
\begin{center}
\includegraphics[width=10cm]{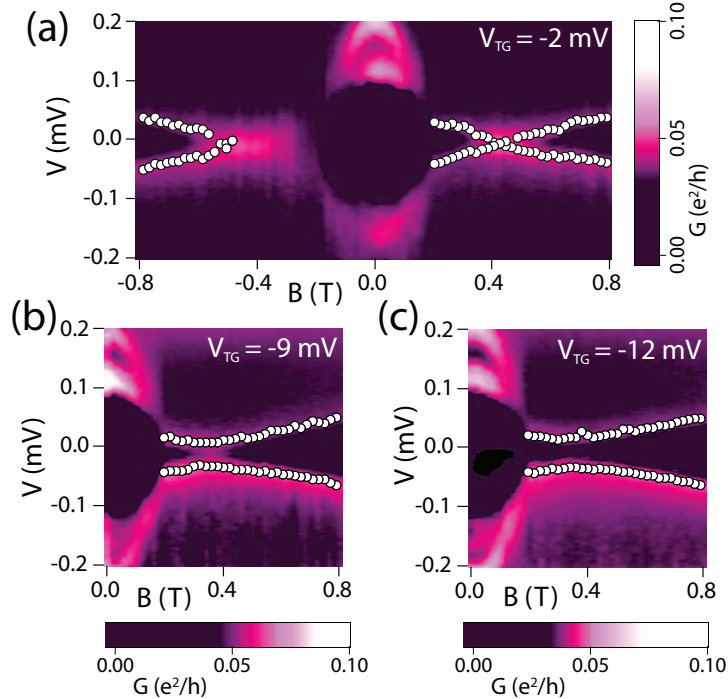}
\caption{Response of the Kondo ridge splitting to $B$ for different
values of ($V_{TG}$,$V_{BG}$). A linear background is subtracted
from the data for better visibility. White dots correspond to maxima
in $G(V)$. \label{fig4}}
\end{center}
\end{figure}

To gain more insight into the origin of the splitting observed in
figure~\ref{fig3}, we have measured the nonlinear conductance $G(V)$
as a function of magnetic field $B$ for three fixed gate
configurations, corresponding to three different points on the
dashed line in figure~\ref{fig2}b. In figure~\ref{fig4}a, we present
data for ($V_{TG} = -2$~mV, $V_{BG} = -0.471$~V). At low field,
superconductivity dominates electrical transport, preventing the
appearance of the Kondo effect. For $\mid B\mid > 0.2$~T,
superconductivity is suppressed, and we can follow the evolution of
the two spin components of the Kondo ridge in the external magnetic
field. From the slope of these lines, we estimate $g \sim 2.24 \pm
0.1$, a value that is consistent with a free electron g-factor plus
some small orbital magnetic moment contribution that could be
present \cite{Minot04}, as the magnetic field was applied under an
angle of $\approx 45^o$ to the axis of the nanotube. The Kondo
resonance is recovered only at a finite magnetic field of $B =
0.44$~T. Note that except for a very small shift of $ 0.007$~T, that
is probably due to flux trapping in the superconducting magnet, the
data for positive and negative magnetic field are equivalent.
Extrapolating the high conductance lines towards $B = 0$, we find a
zero field splitting of $\Delta E \approx 120~\mu eV$.

A qualitatively different picture evolves at two other positions in
the $V_{TG}/V_{BG}$-plane of figure~\ref{fig2}b : For
(-9~mV/-0.456~V) (figure~\ref{fig4}b) and (-12~mV/-0.453~V)
(figure~\ref{fig4}c), the Kondo resonance is not recovered for any
value of magnetic field, but instead the two high conductance lines
only approach each other up to a minimum distance of about
$50~\mu$V, giving an anti-crossing like behavior. Note that both of
these points correspond to the middle of the same charge state as
the measurement in figure~\ref{fig4}a. Extrapolating to $B = 0$
gives roughly the same zero field splitting $\Delta E \approx
120~\mu$eV, and the slope of the two high conductance lines, taken
at higher magnetic fields, is the same as for the data in
figure~\ref{fig4}a.

At first sight our data bear a striking resemblance to data obtained
on a Carbon nanotube quantum dot with ferromagnetic contacts
\cite{Hauptmann08}. In this experiment, a splitting of the Kondo
resonance at $B = 0$ as well as its recovery at finite magnetic
field could be explained with the presence of an exchange field,
that is induced by the magnetization of the ferromagnetic electrodes
\cite{Martinek03,Martinek05}. An avoided crossing similar to the one
in figures~\ref{fig4}b/c had been observed in some samples in
\cite{Hauptmann08}, and was explained with a small misalignment of
the exchange field with respect to the external magnetic field, a
situation which would expose the impurity spin to a finite magnetic
field for all values of the external magnetic field, and therefore
prevent the recovery of the Kondo resonance. This misalignment is
however given by the domain structure of the ferromagnetic
electrodes, and cannot be changed by gate voltages. Moreover, the
existing theories for a ferromagnetic exchange field acting on a
single spin in a quantum dot \cite{Martinek03,Martinek05} predict
the induced exchangefield to always be parallel (or anti-parallel)
to the magnetization of the electrodes, a misalignment of the two is
not possible. Additionally, we have not deliberately introduced any
ferromagnetic parts into our sample, and therefore have to discard
an explanation in terms of a ferromagnetic exchange field.

There have been reports of a split Kondo effect in CNT quantum dots
\cite{Nygard04}, where the coupling to a ferromagnetic catalyst
particle, that is needed to synthesize the Carbon nanotube
\cite{Kong98}, has been discussed as the origin of a ferromagnetic
exchange field that could split the Kondo resonance at $B = 0$.
However, as the catalyst particles normally stick to the end of the
nanotube (at several $\mu$m distance from the quantum dot), and are
extremely small (1-2~nm, equivalent to the nanotube diameter), we do
not think that they can account for an exchange field on the dot of
the order of $0.4$~T. We believe that the nonlinear temperature
dependence of the equilibrium conductance at $B = 0$ reported in
\cite{Nygard04} rather points to a two-impurity Kondo effect as the
origin of the observed splitting of the Kondo resonance.

The Kondo effect in the presence of two magnetic impurities has been
investigated first theoretically \cite{Jayaprakash81,
Tolea07,Chang09} and later experimentally \cite{vanderWiel02,
Craig04}. In quantum dots, a two-impurity Kondo effect can occur in
a parallel double quantum dot or in a single dot with two
near-degenerate levels. Two electrons (impurities) that can interact
with each other via exchange interaction, and that interact
independently with the Fermi sea in the leads, are considered.
Depending on temperature, magnetic field, strength and sign of the
inter-impurity exchange interaction, the occurrence of the Kondo
effect is qualitatively different from the simple spin-$1/2$ Kondo
effect. Here we will consider a situation where one of the
impurities is fixed, and doesn't participate in transport, e. g. an
electron localized in a trap state in the gate oxide close to the
nanotube. However, it will still interact with the spin of the
second impurity via some exchange interaction
\begin{equation}
E_{ex} =  J \quad \mathbf{S_1} \cdot \mathbf{S_2}.
\end{equation}
Such a situation has been realized in a serial triple quantum dot
device \cite{Craig04}, where the two outer dots held the impurity
spins, and the larger middle dot functioned as a small Fermi
reservoir, mediating a tunable Ruderman-Kittel-Kasuya-Yoshida (RKKY)
like exchange interaction. This
experiment has been described theoretically in Refs.
\cite{Simon05,Vavilov05}. Common to both ferromagnetic and
antiferromagnetic exchange interaction between the two impurities is
a suppression of conductance at $V = 0$ for zero magnetic field,
which results in a split Kondo resonance. To distinguish between
ferro- and antiferromagnetic exchange between the two electrons, one
has to look at the magnetic field dependence
\cite{Simon05,Pustilnik01,Chung07,Aono01}, as only for
antiferromagnetic exchange ($J>0$) the two peaks approach each other
with increasing magnetic field, and the Kondo resonance can be
recovered at finite B. In case of ferromagnetic exchange ($J<0$),
the magnetic field dependence of the two maxima in $G(V)$ is much
weaker, and a more complicated two stage Kondo effect will occur at
low enough temperature. In the case of $J>0$, the antiferromagnetic
interaction between the two unpaired spins is competing with the
Kondo screening, the two impurities are locked into a singlet, and
the Kondo effect is inactive, which leads to a suppression of
conductance at $V = 0$. Only for larger energies, when transition
from the singlet into the lowest triplet state can take place, the
Kondo screening will be reestablished and two maxima in $G(V)$ at $V
= \pm J/2$ will occur \cite{Simon05}. As the two maxima in this case
correspond to a singlet/triplet degeneracy point, they will change
their position with magnetic field and allow for a recovery of the
Kondo resonance at the magnetic field induced crossing between the
singlet and lowest triplet state.

With this scenario in mind, we can explain the data shown in
figure~\ref{fig4}a. The observed zero-field splitting of $\Delta E
\approx 120~\mu$eV would correspond to the exchange constant between
the electron spin and an additional spin, that is fixed, either in a
second dot somewhere on the nanotube, or at a defect in the gate
oxide close to the nanotube. An exchange constant $J$ of the order
of $100~\mu$eV is realistic \cite{Craig04}. The approaching of the
two maxima and the recovery of the Kondo resonance at $B = 0.44$~T
also correspond to the theoretical prediction, and are qualitatively
similar to experimental results reported previously
\cite{Heersche06,Osorio07,Otte09}. The effects of a fixed impurity
spin have also been investigated in two recent experiments on Carbon
nanotube quantum dots \cite{Bomze10,Chorley10}.

To explain the data of figures~\ref{fig4}b/c in this scenario, we
have to extend the inter-impurity exchange interaction by an
antisymmetric, Dzyaloshinskii-Moriya (DM) type component
\begin{equation}
E_{Ex}' = J \quad \mathbf{S_1} \cdot \mathbf{S_2} + {\mathbf D}
\cdot \mathbf{S_1} \times \mathbf{S_2}
\end{equation}
\cite{Dzyaloshinskii58,Moriya60}, which would prevent the complete
anti-parallel alignment of the two spins, having the same effect as
a misaligned exchange field in a dot with ferromagnetic contacts
\cite{Hauptmann08}. As the presence of a second, fixed spin, and the
exchange interaction of this fixed spin with the dot spin in our
sample is not well controlled and rather accidental, such a
situation is imaginable. Similar to the exchange constant $J$, the
vector ${\mathbf D}$ is a property of the two-spin system and in the
case of the DM interaction is determined by spin-orbit interaction.
It is therefore possible that the direction and length of ${\mathbf
D}$ are sensitive to the electrostatic environment of the two spins,
so that $E_{Ex}'$ can be gate dependent. To our knowledge, there are
no previous reports of antisymmetric exchange interactions in
quantum dots, although there is some theoretical work on electronic
transport through a molecular magnet that is composed of two spins
with DM interaction \cite{Herzog10}. This work however examines the
level spectrum of the molecular magnet as observed from excited
state lines in sequential tunneling through a weakly coupled quantum
dot. A two-impurity Kondo effect with spin-orbit coupled conduction
electrons has been investigated in \cite{Mross09}, and the Kondo
effect in a quantum dot with an even number of electrons in the
presence of
 spin-orbit interaction has been investigated in \cite{Lucignano10}.
However, we are however not aware of any theoretical work that
corresponds to our experimental situation with an odd number of
electrons on the dot and on-site spin-orbit interaction.

To support the explanation in terms of a two-impurity Kondo effect,
a detailed determination of the temperature dependent conductance
would be desirable. Unfortunately, due to the bad electrostatic
stability on longer timescales, this has not been possible for the
present device. We note, however, that the non-monotonic temperature
dependence observed in \cite{Nygard04} corresponds to the prediction
for a two-impurity Kondo effect, as given e. g. in \cite{Simon05}.

In conclusion, we have shown a split Kondo effect in a Carbon
nanotube quantum dot with additional, locally acting gates. The
zero-bias Kondo resonance can be recovered at a finite magnetic
field, a situation consistent with a two-impurity Kondo effect of
two spin 1/2 particles that are linked by an antiferromagnetic
exchange interaction. A slight deformation of the dot confinement
potential by one of the local gates led to a situation where the
Kondo resonance could no longer be recovered, but where the two high
conductance lines form an anti-crossing, with a minimum distance of
about $50~\mu$eV. We speculate that this anti-crossing might be due
to the occurrence of a small antisymmetric component in the exchange
interaction between the two Kondo impurities, which becomes active
for certain gate voltages and prevents a complete compensation of
the impurity spin by the Fermi sea in the leads. Further
experimental and theoretical work will be needed to unambiguously
support this picture.

\section*{Acknowledgements}

Financial support from the EU-FP6 project HYSWITCH, the Swiss
National Science Foundation, and the NCCR on Nanoscale Science is
acknowledged. M. W. is a grantee of the Treubel Fund. We thank
Pascal Simon and Mahn-Soo Choi for fruitful discussion.

\end{document}